\documentclass[pra,aps,groupedaddress,floatfix,twocolumn,superscriptaddress,showpacs]{revtex4-1}
\usepackage{amsmath,amssymb,epsfig,bm,pifont}
\usepackage[linkcolor=black,citecolor=black,urlcolor=blue,colorlinks=true,linktocpage=true]{hyperref}
\usepackage{graphicx}
\usepackage{epstopdf}
\usepackage{bm}
\usepackage{tabularx}

\def\0#1#2{\frac{#1}{#2}}

\def\s0#1#2{\mbox{\small{$ \frac{#1}{#2} $}}}

\def\CZ{{\mathcal Z}}
\def\CD{{\mathcal D}}

\newcommand{\beq}{\begin{equation}}
\newcommand{\eeq}{\end{equation}}
\newcommand{\bea}{\begin{eqnarray}}
\newcommand{\eea}{\end{eqnarray}}
\newcommand{\tr}{\mathrm{tr}}

\begin{document}

\title{Ground-state energy, density profiles, and momentum distribution of attractively interacting 1D Fermi gases
with hard-wall boundaries: a Monte Carlo study}

\author{J. R. McKenney}
\email{joshmck@unc.edu}
\affiliation{Department of Physics and Astronomy, University of North Carolina, Chapel Hill, NC, 27599, USA}

\author{C. R. Shill}
\email{crshill@live.unc.edu}
\affiliation{Department of Physics and Astronomy, University of North Carolina, Chapel Hill, NC, 27599, USA}

\author{W. J. Porter}
\email{wjporter@live.unc.edu}
\affiliation{Department of Physics and Astronomy, University of North Carolina, Chapel Hill, NC, 27599, USA}

\author{J. E. Drut}
\email{drut@email.unc.edu}
\affiliation{Department of Physics and Astronomy, University of North Carolina, Chapel Hill, NC, 27599, USA}

\begin{abstract}
Motivated by the realization of hard-wall boundary conditions in experiments with ultracold atoms, 
we investigate the ground-state properties of spin-1/2 fermions with attractive interactions in a one-dimensional box.
We use lattice Monte Carlo methods to determine essential quantities like the energy, which we compute as a function
of coupling strength and particle number in the regime from few to many particles. Many-fermion systems bound by hard walls 
display non-trivial density profiles characterized by so-called Friedel oscillations (which are similar to those observed in harmonic 
traps). In non-interacting systems, the characteristic length scale of the oscillations is set by $(2 k^{}_F)^{-1}$, where $k^{}_F$ is the 
Fermi momentum, while repulsive interactions tend to generate Wigner-crystal oscillations of period $(4 k^{}_F)^{-1}$.
Based on the non-interacting result, we find a remarkably simple parametrization of the density profiles of the attractively interacting case, which
we generalize to the one-body density matrix. While the total momentum is not a conserved quantity in the presence of hard walls,
the magnitude of the momentum does provide a good quantum number. We are therefore able to provide a detailed
characterization of the (quasi-)momentum distribution, which displays rather robust discontinuity at the Fermi surface.
In addition, we determine the spatially varying on-site density-density correlation, which in turn yields Tan's contact density and, 
upon integration, Tan's contact. As is well known, the latter fully determines the short-range correlations and plays a crucial role in 
a multitude of equilibrium and non-equilibrium sum rules.
\end{abstract}

\date{\today}
\maketitle

\section{Introduction} 

As is well-known, advances in trapping, cooling, and
manipulation of ultracold atoms, single species and mixtures alike, have made it possible
to realize controlled studies of quantum systems in strongly coupled few- and many-body 
regimes~\cite{ReviewsAtoms}.
The already large set of possibilities allowed by those experimental techniques continues 
to expand further and faster than ever.
Understanding strongly coupled quantum matter in these atomic systems, in the 
wide variety of available scenarios, is of broad interest: strongly correlated matter is a challenging
problem that pervades all energy scales, from quantum chromodynamics to condensed 
matter physics.

One of the most fruitful exchanges has actually been with the area of nuclear and neutron matter structure.
In the latter, interactions are of limited range, and scattering lengths are comparatively large, a situation that
bears strong similarities with ultracold atomic fermions close to a Feshbach resonance, differences
notwithstanding~\cite{ZinnerJensen}. 
Similarly, the realization of periodic systems via optical lattices~\cite{OpticalLattices} and the creative approach to 
engineering topologically non-trivial phases using internal degrees of freedom~\cite{TopoNature} have recently 
strengthened the connections to condensed matter physics.

New experimental techniques allow for the implementation of (quasi-) hard-wall traps. This development is interesting because
the ``flat-bottom'' of the trap simulates a uniform system~\cite{Hadzibabic}. Close to the boundaries, or for low particle numbers,
the edge effects result in the so-called Friedel oscillations, i.e. deviations from uniformity.
The original work by Friedel actually considered the problem of the density variations due to the presence of an impurity~\cite{Friedel}
in an electronic system. Since then, other authors considered similar ``impurity'' problems (see e.g.~\cite{Tutto, Fernandez, Chaterjee}), 
and more recently the application of Friedel oscillations as probes for quasiparticles has been advocated~\cite{Demler}. 

In addition to the above, there are other motivations to understand Friedel oscillations in detail.
For instance, it was shown by Zhang et al. in Ref.~\cite{ZhangEtAl} 
that the presence of boundaries and interfaces in nanocomposites and heterostructures limits their mechanical strength 
precisely due to the appearance of Friedel oscillations. Clark et al.~\cite{ClarkEtAl} 
showed that Friedel oscillations near a bilayer-monolayer graphene interface open a gap at the Fermi energy (for electrons with 
wave vectors normal to said interface), which consequently affects transport properties across the boundary. 
The interplay between the shape of the Fermi surface and Friedel oscillations (in particular for an oxygen impurity on 
the surface of a ferromagnetic thin film) was studied by Bouhassoune et al.~\cite{BouhassouneEtAl},
where ``giant'' directional effects were reported and shown to be tunable, with consequences for nanospintronics applications.
Generally speaking, Friedel oscillations continue to be of interest to solid-state physics and materials science, even though their 
existence has been qualitatively understood for a while.

In this work, we do not consider impurities per se, but treat the problem 
of density oscillations resulting from ``open'' or ``hard-wall'' boundary conditions in a non-relativistic system of spin-$1/2$ fermions.
Previous work studied this problem in the electronic case using bosonization~\cite{Fabrizio, Grabert} or a variety of numerical 
methods~\cite{Bedurftig,Capelle,Xianlong,XuEtAl,Angelone}.
Using bosonization and density-matrix renormalization group (DMRG) techniques, 
Ref.~\cite{Eggert} studied a related situation, namely the {\it repulsive} Hubbard model 
in 1D with hard-wall boundaries. It was then found that the repulsive interaction parametrizes a crossover
between Friedel-type density oscillations (characterized by a $2k^{}_F$ wavelength) at weak coupling and
Wigner-crystal-type oscillations (with $4k^{}_F$ wavelength) at strong coupling. In the latter regime, the repulsion
forces particles of opposite spins to occupy higher momentum states, effectively doubling the Fermi momentum.

As mentioned above, here we quantitatively explore several properties of spin-$1/2$ fermions in 1D with {\it attractive} short-range interactions 
confined by hard-wall boundaries. We calculate density profiles and study the enhancement of Friedel oscillations
with increasing interaction strength; determine the occupation of standing-wave orbitals; and find the ground-state energy.
In addition, we present a detailed characterization of the one-body density matrix and the short-distance correlations by computing the so-called contact density~\cite{Tan},
which in these systems is a spatially varying quantity. As is well known, the contact fully characterizes the short-distance
behavior of correlations and enters sum rules of transport coefficients (see e.g.~\cite{ContactReview}).
We study all of the above in the few- to many-body range of $N=8,..., 24$ unpolarized particles and cover weakly to 
strongly coupled regimes, as measured by the conventional dimensionless coupling $\gamma$ (see below).

Naturally, our work complements Ref.~\cite{Eggert} mentioned above in that we study the regime of attractive interactions, 
but we also expand on that work by considering a range of particle numbers. In contrast to the repulsive 
case, strong pairing correlations induced by attractive interactions lead to two-body bound-state formation and an enhancement 
of the Friedel oscillations, with a characteristic coupling-dependent discontinuity in the momentum distribution.
The effect on Friedel oscillations was also noted in Ref.~\cite{BAD}, where a similar calculation was carried out
in the presence of a harmonic trapping potential (see also Ref.~\cite{NJP2015}).

Our calculations are also a first step towards exploring the properties of analogue systems in
higher dimensions. As outlined in Ref.~\cite{BDP}, this is enabled by Fourier acceleration, which
can be adapted to hard-wall boundaries without affecting the dominant scaling of computational
cost with system size. The same methods can be applied to systems in optical lattices, in particular
in cases of internal degeneracies $N_f^{} > 2$.

Generally speaking, correlations tend to be stronger in 1D systems than in their 2D and 3D counterparts~\cite{GiamarchiBook}.
For this reason, investigating and implementing non-perturbative methods [such as exact diagonalization, quantum Monte Carlo (QMC), 
density-matrix renormalization group (DMRG), Bethe ansatz (BA), bosonization, etc.] are essential.
However, not all of these methods are universally applicable. 
For instance, quantum Monte Carlo can address problems in any number of spatial dimensions, but suffers from a sign 
problem at finite polarization and for repulsive interactions.
The BA can solve the problem exactly for attractive or repulsive short-range interactions in 1D, and it has been widely applied  
for infinite systems (see e.g. Ref.~\cite{BAReview}). The BA is not applicable in higher dimensions
or in the presence of external trapping potentials, as translation invariance is broken, but it can and has been applied 
to 1D infinite square-well traps with repulsive interactions~\cite{BAwHWBC}. The DMRG method has also been used to
analyze that system~\cite{Angelone}.
Finally, bosonization methods are often restricted to very low temperatures where the dispersion relation around the 
Fermi points can be well approximated as linear. As argued in Ref.~\cite{BeyondBosRev}, however, it has become
clear in the last few years that considering non-linearities is important, as they account for characteristic behavior even at the level of broad, qualitative features, most 
noticeably so in the system's dynamic response functions~\cite{LLApprox}.
For these reasons, there have recently been a number of exact diagonalization (and similar) studies, with focus on
pairing correlations in spin-polarized, harmonically trapped systems (see e.g. Refs.~\cite{DAmicoRontani, LindgrenEtAl, SowinskiEPL, SowinskiFBS}).
The same reasons further motivate the present work.

\section{Many-body method}

\subsection{Hamiltonian and basic formalism}

In this work we will focus on a non-relativistic Hamiltonian $\hat H$, with short-range interactions,
as a model relevant for matter in dilute regimes. Thus,
\beq
\hat H = \hat T + \hat V,
\eeq
where 
\beq
\hat{T} = \sum_{s=\uparrow,\downarrow}{\int_0^{L}{d x\,\hat{\psi}^{\dagger}_{s}(x)\left(-\frac{\hbar^2\partial^2_x}{2m}\right)\hat{\psi}^{}_{s}(x)}}
\eeq
is the kinetic energy, and
\beq
\label{Eq:InteractionV}
\hat{V} = -g\int_0^{L}{dx\,\hat{n}^{}_{\uparrow}(x)\hat{n}^{}_{\downarrow}(x)},
\eeq
is the two-body, zero-range interaction. Although we have written $\hbar$ and $m$ explicitly above,
we will take them to unity from this point on and similarly for Boltzmann's constant $k_B$.
While $g$ is the bare coupling, it is physically meaningful in 1D: 
$g = 2/a_0$, where $a_0$ is the scattering length. As conventional in studies of uniform systems,
we define a dimensionless coupling
\beq
\gamma\equiv \frac{g L}{N},
\eeq
where $L$ is the physical extent of the system and $N$ is the total particle number.

As in previous work~\cite{GSC1D}, we approach the ground state of our many-body quantum system 
by imaginary-time evolution of a ``guess'' state $| \psi^{}_0 \rangle$
\beq
| \psi \rangle_\beta^{} = e^{-\beta \hat{H}} \,| \psi^{}_0 \rangle,
\eeq
where $\beta$ is the extent of the imaginary time evolution.
This approach is feasible if the guess has a non-vanishing projection onto the true ground state.
We take $|\psi^{}_0\rangle$ to be a Slater determinant of single-particle orbitals $\{\varphi^{}_k\}$ given by plane waves with hard-wall boundary conditions, i.e. 
\beq
\varphi_k(x) = \sqrt{\frac{2}{L}} \sin\left( \frac{\pi k x}{L}\right),
\eeq
where $L$ is the size of the box, and $k$ is a positive integer.
More specifically, we take $k = 1,2,\dots, N^{}_\uparrow$ with $N^{}_\uparrow = N^{}_\downarrow = N/2$ being the number of fermions of each species.
In this method, it is essential to have access to the operator
\beq
\hat{U}(t',t) \equiv \exp\left[-(t'-t)\hat{H}\right],
\eeq
which is a complicated object in general, but which can be approximated using a 
Suzuki-Trotter decomposition
\beq
\label{Eq:Ut}
\hat{U}(t+\tau,t) =e^{-\tau\hat{T}/2}e^{-\tau\hat{V}}e^{-\tau\hat{T}/2}+O(\tau^3),
\eeq
where $\tau$ is our imaginary-time discretization parameter.
This factorization allows us to use, at each time step $t$, an auxiliary-field Hubbard-Stratonovich 
transformation of the interaction, i.e. a representation of the two-body interaction via one-body potentials:
\beq
e^{-\tau\hat{V}} = \int \CD \sigma(x)\,
e^{-\tau \hat{V}^{}_{\uparrow,\sigma}}e^{-\tau \hat{V}^{}_{\downarrow,\sigma}},
\eeq
where the $\hat{V}^{}_{s,\sigma}$ are external-potential, one-body operators that depend on the
auxiliary field $\sigma(x)$. The integral $\int \CD \sigma(x)$ is a sum
over all possible configurations of $\sigma$ at the specific time $t$ [see Eq.~(\ref{Eq:Ut})].

Combining the above steps, one finds
\beq
\hat{U}(\beta,0) = \int \CD \sigma (x,t) \prod_t \hat{U}_\sigma(t+\tau,t),
\eeq
where the path integral is over all possible spacetime dependent fields $\sigma (x,t)$ and
the $\sigma$-specific evolution operator is
\beq
\hat{U}_\sigma(t+\tau,t) \equiv
e^{-\tau\hat{T}/2}e^{-\tau\hat{V}_{\uparrow,\sigma}}e^{-\tau\hat{V}_{\downarrow, \sigma}}e^{-\tau\hat{T}/2},
\eeq
which is a product of exponentials of one-body operators. 

Moreover, we identify a zero-temperature partition sum
\beq
\label{Eq:Z}
\CZ \equiv {\langle \psi^{}_0 | \, \hat{U}(\beta,0) \,| \psi^{}_0 \rangle} = \int \CD \sigma (x,t)\, P[\sigma],
\eeq
where we defined
\beq
\label{Eq:PsigmaDef}
P[\sigma] \equiv \langle \psi^{}_0|\,\hat{U}_\sigma(\beta,0)\, |\psi^{}_0 \rangle .
\eeq
Since $\hat U_\sigma$ is composed of a string of exponentials of one-body operators, a well-known 
result of second-quantization formalism indicates that (assuming our guess state $ |\psi^{}_0 \rangle$ is a 
single Slater-determinant state, as specified above)
\beq
\label{Eq:Psigma}
P[\sigma] = \langle \psi^{}_0| \,\hat{U}_\sigma(\beta,0)\, |\psi^{}_0 \rangle = {\det}^2 \left[{M}_\sigma(\beta) \right],
\eeq
where the power of two results from our system having two distinguishable (but otherwise identical)
fermion species. The matrix ${M}_\sigma(\beta)$ is the one-particle representation of
$\hat{U}_\sigma(\beta,0)$, restricted to the Hilbert space of the occupied orbitals, i.e.,
\beq
\label{Eq:MDef}
\left[ {M}_\sigma(\beta) \right]^{}_{ab}  = \langle a | \hat{U}_\sigma(\beta,0) | b \rangle,
\eeq
where $a,b = 1,2,\dots, N^{}_\uparrow$.
Using this auxiliary field formalism, one may sample $\sigma$ according
to $P[\sigma]$ using well-known methods. For the calculations carried out in this work, 
we used the lattice hybrid Monte Carlo algorithm~\cite{reviewMC,HMC}.

\subsection{Expectation values of operators}

The above formalism enables the non-perturbative evaluation of arbitrary observables,
as long as $P[\sigma]$ is non-negative (which in our case it is, as mentioned above),
as we shall see next.
The ground-state expectation value of an operator $\hat O$ is
\beq
\langle \hat O \rangle = \lim_{\beta\to\infty} O_\beta,
\eeq
where
\beq
\label{Eq:ExpVal}
O_\beta \equiv \frac{\langle \psi^{}_0 |
\,\hat{U}(\beta,\beta/2)\,\hat{O}\,\hat{U}(\beta/2,0)\,| \psi^{}_0 \rangle}
{\langle \psi^{}_0 | \, \hat{U}(\beta,0) \,| \psi^{}_0 \rangle}.
\eeq

\begin{figure}[b]
\includegraphics[width=1.0\columnwidth]{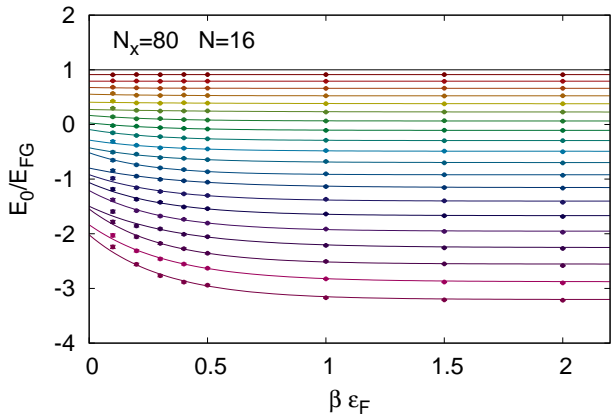}
\caption{\label{Fig:EnergyRainbowBeta} 
Convergence of our energy estimator as a function of 
$\beta \varepsilon^{}_{F}$, for $N=16$ particles in a 1D segment (discretized using $N_x^{}=80$ points), 
for couplings $\gamma=0.0, 0.2, ..., 4.0$ (top to bottom).
}
\end{figure}

Once the time-evolution operators $\hat U$ are written in field-integral form, as shown in the 
previous subsection, Eq.~(\ref{Eq:ExpVal}) becomes
\beq
\label{Eq:ExpValPathIntegral}
O_\beta = \frac{1}{\CZ}\int \CD \sigma\, P[\sigma] \, O[\sigma],
\eeq
where
\beq
O[\sigma] \equiv \frac{\langle \psi^{}_0 |
\,\hat{U}_\sigma(\beta,\beta/2)\,\hat{O}\,\hat{U}_\sigma(\beta/2,0)\,| \psi^{}_0
\rangle}{\langle \psi^{}_0 | \, \hat{U}_\sigma(\beta,0) \,| \psi^{}_0 \rangle}.
\eeq
This path integral form Eq.~\ref{Eq:ExpValPathIntegral} is a function of the imaginary time $\beta$, which should approach the 
ground-state answer when extrapolated to large $\beta$ (see Fig.~\ref{Fig:EnergyRainbowBeta}).

%
%

In practice, the calculation of $O[\sigma]$ above, for a given configuration of the field $\sigma$, can be easily carried out when 
$\hat{O}$ is generic one-body operator, in the following fashion. Going back to Eqs.~(\ref{Eq:Z})-(\ref{Eq:MDef}), we 
insert a source factor $\exp{(\lambda \hat O)}$ as follows; we take
\beq
 \langle \psi^{}_0| \,\hat{U}_\sigma(\beta,0)\, |\psi^{}_0 \rangle \to 
  \langle \psi^{}_0| \,\hat{U}_\sigma(\beta,\beta/2) e^{\lambda \hat O}\hat{U}_\sigma(\beta/2,0)\, |\psi^{}_0 \rangle
\eeq
Since the source factor is the exponential of a one-body operator, it modifies the form of $M_\sigma$ defined above
in a predictable way.
Inserting the resulting expression in Eq.~(\ref{Eq:Z}) (or rather its natural logarithm), differentiating with respect to $\lambda$, 
and taking $\lambda=0$, it is easy to identify
\beq
\label{Eq:Observable}
O[\sigma] = \tr[{M}_\sigma^{-1}{U}_\sigma(\beta,\beta/2)\,{O}\,{U}_\sigma(\beta/2,0)],
\eeq
where the trace is over the space of occupied orbitals, and ${U}_\sigma(t,t')$ and ${O}$ are the
single-particle matrix representations of $\hat {U}_\sigma(t,t')$ and $\hat{O}$, respectively. Note, in particular, 
that the matrix product of ${U}_\sigma$ and ${O}$ above is over the full single-particle space, not just the occupied 
orbitals.

\section{Results}

Using the formalism presented above, we carried out lattice calculations in fixed system sizes of length $L=(N^{}_x+1)\ell$,
discretized by setting $N_x^{}=20,30,40,60,80$ and maintaining $L$ constant. Below we present plots for several quantities for $N_x^{}=80$,
and discuss finite-size effects in a later section. The extent of the time direction, as measured by 
the dimensionless parameter $\beta \varepsilon^{}_{F}$, was varied so as to allow for a meaningful extrapolation to the ground state,
as explained below. Here, $\varepsilon^{}_{F} = k_F^{2}/2$, and $k_F=\pi N / (2 L)$, where $N$ is the total particle number.
In such lattices, we studied systems of $N=8,12,...,24$ particles by
taking $10^4$ samples of the auxiliary field, which yields statistical uncertainties on the
order of $1\%$. Finally, we varied the strength of the interaction between $\gamma=0$ and $\gamma=4.0$.
\begin{figure}[t]
\includegraphics[width=1.0\columnwidth]{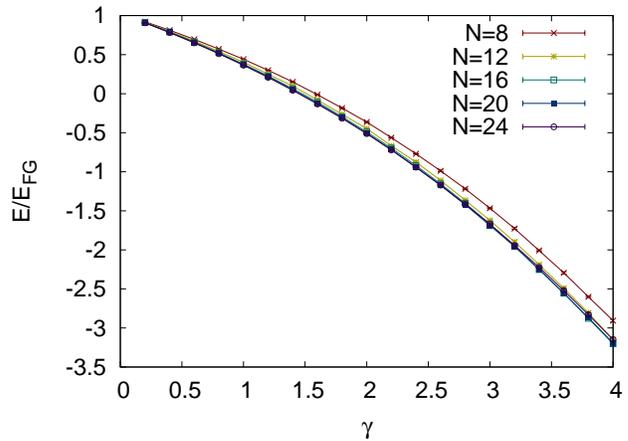}
\caption{\label{Fig:EnergyGS} 
The ground-energy of $N=8,...,24$ unpolarized fermions in a 1D segment (discretized using $N_x^{} = 80$ points), 
in units of its non-interacting counterpart $E_\text{FG}^{}$, as a function of the dimensionless coupling $\gamma$.
}
\end{figure}
%

\subsection{Ground-state energy}

In order to determine the ground-state energy we used the formalism presented in the previous section, but
used a shortcut: the $\beta$ derivative of the path-integral form of $\ln \mathcal Z$ produces the desired expression, 
up to a constant. It is not difficult to show (see Ref.~\cite{reviewMC}) that the resulting estimator $E_\beta^{}$
approaches the ground-state result $E_\text{GS}^{} $ exponentially, i.e.
\beq
E_\beta^{} \equiv -\frac{\partial \ln \mathcal Z}{\partial \beta} \to E_\text{GS}^{} + K e^{-\beta \Delta},
\eeq
where $K$ is a constant and $\Delta$ is the difference between the energy of the first excited state and the ground state.
In Fig.~\ref{Fig:EnergyRainbowBeta} we show the above exponential fits to Monte Carlo data for a representative case 
($N_x^{}=80$, $N=16$) for several couplings.

In Fig.~\ref{Fig:EnergyGS}, we show our results for the ground-state energy $E_\text{GS}^{}$ in units of the 
energy of the non-interacting case
\beq
E_\text{FG}^{} = \frac{\pi^2}{L^2} \sum_{k=1}^{N/2} k^2 = \frac{\pi^2}{24 L^2}N(N+1)(N+2).
\eeq
As is evident from the figure, for all the couplings we studied, the ground-state energy appears to approach the
large-$N$ limit very quickly.

\subsection{Density profiles}

%
\begin{figure}[t]
\includegraphics[width=1.0\columnwidth]{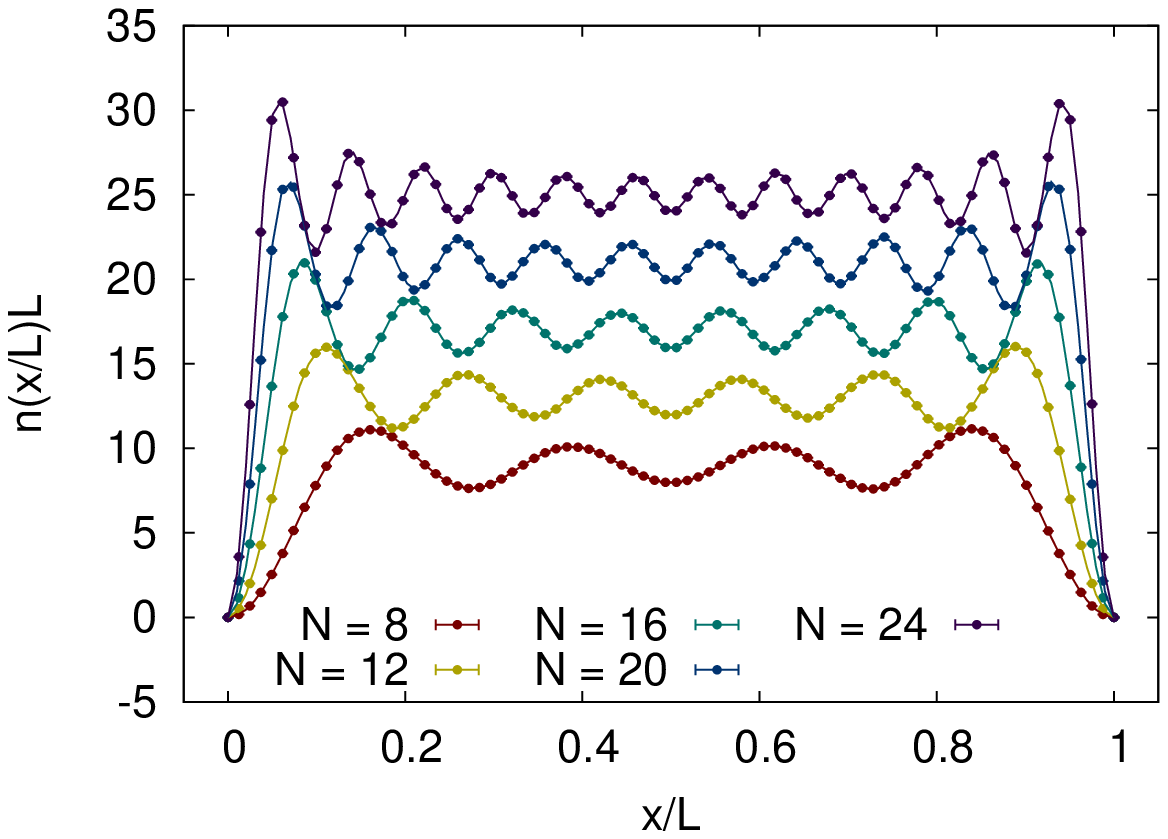}
\includegraphics[width=1.0\columnwidth]{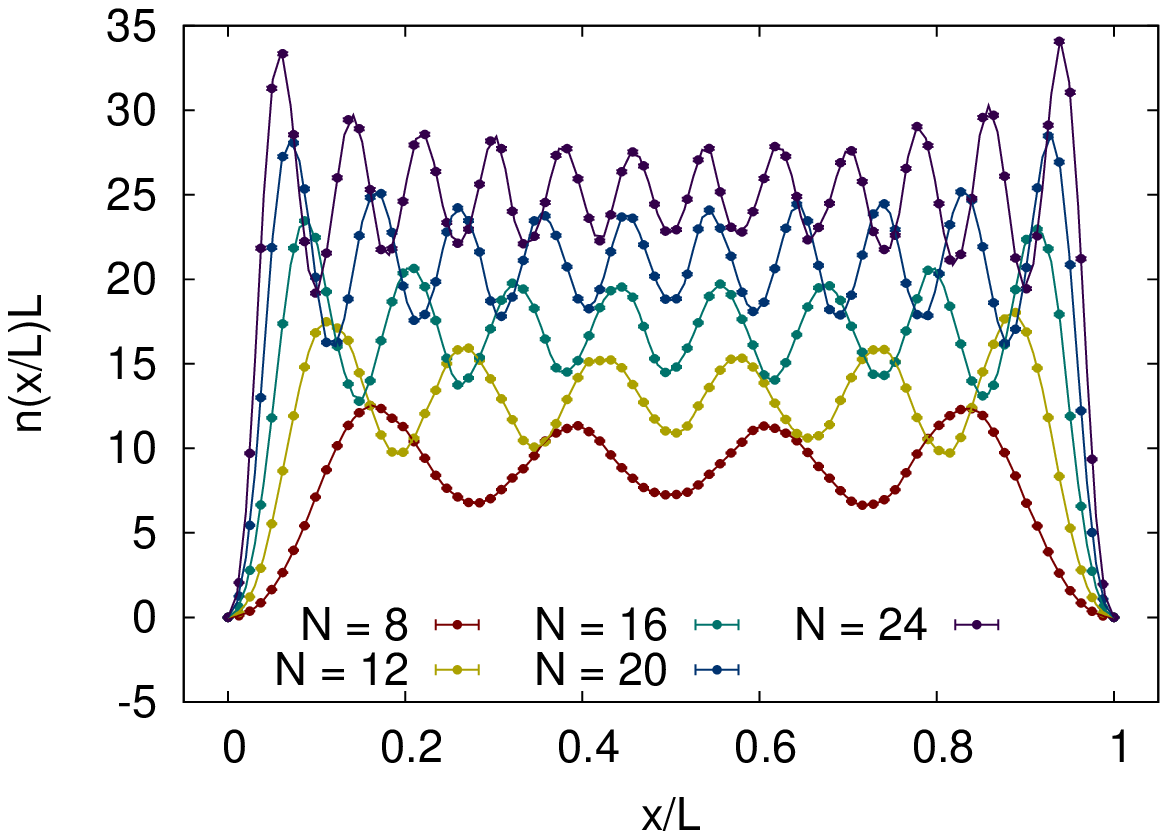}
\caption{\label{Fig:DensProfCoupling}  Density profiles versus the scaled position $x/L$
for $N^{}_{x}=80$ at weak coupling ($\gamma=0.2$, top panel) and strong coupling ($\gamma=3.0$, bottom panel), 
for particle numbers $N=8,12,16,20,24$ (from bottom to top).
}
\end{figure}
%
%
\begin{figure}[t]
\includegraphics[width=1.0\columnwidth]{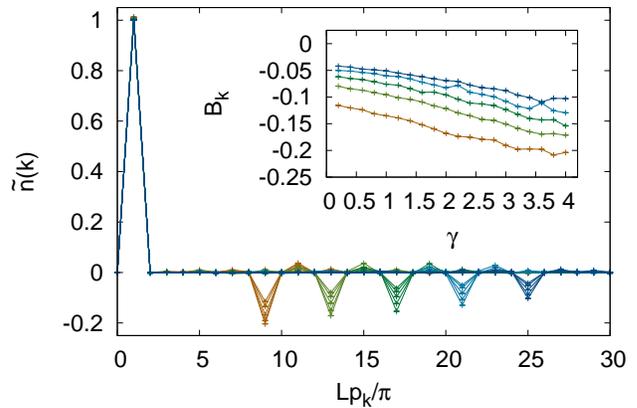}
\caption{\label{Fig:SineTransf}  Hard-wall transform coefficients $B_k$ of Eq.~(\ref{Eq:SineTransf}), 
as a function of $L p_k/\pi = k$, for $N_x^{}=80$ and several values of the coupling $\gamma$ and particle number $N$. 
Note the maximum at $k=1$ is equal to unity to an accuracy better than 2\% for all $\gamma$ and $N$; the minima, 
on the other hand, show clear variation with increasing $N=8,...,24$ (left to right) as well as increasing $\gamma$ (top to bottom; see inset).
Inset: Value of the minimum as a function of the coupling $\gamma$ for $N=8,...,24$ (bottom to top).
}
\end{figure}
%

To calculate the density profiles we used the expression of Eq.~(\ref{Eq:Observable}), where $\hat O = \hat n(x)$,
which indicates that
\bea
\label{Eq:Density}
n(x) &=& \langle \hat n (x)\rangle = \frac{1}{\mathcal Z} \int \mathcal D \sigma \, P[\sigma] \, n[\sigma, x],
\eea
where
\bea
n[\sigma, x] \equiv
\sum_{a,b,c}^{N/2}
\left[{M}_\sigma^{-1}\right]^{}_{ab}
\left[{U}_\sigma(\beta,\beta/2)\right]^{}_{bx} 
\left[\,{U}_\sigma(\beta/2,0)\right]^{}_{xa}.
\eea
To derive the previous expression, we have used that the single-particle representation of $\hat O = \hat n(x)$ is simply 
$O_{y,y'} = \langle y |\hat n(x) | y' \rangle = \delta(y,y')\delta(x,y)$, where $| y \rangle$ is a coordinate eigenstate.

In Fig.~\ref{Fig:DensProfCoupling} we show the density profiles for two different values of the attractive coupling $\gamma$
as a function of particle number $N$. As is evident from the figure, the main effect of attractive interactions is to enhance the 
amplitude of the Friedel oscillations while their frequencies are maintained. In order to analyze the
oscillations in more detail, we recall that the density profile of the noninteracting case can be easily obtained by summing the contributions
of the $N/2$ occupied single-particle states, namely
\bea
\label{Eq:nxNonInt}
n(x/L) &=& \frac{2}{L} \sum_{k=1}^{N/2} \sin^{2}\left(\frac{\pi k x}{L}\right) \nonumber \\
&=& \frac{N+1}{L} - \frac{1}{L}\frac{\sin\left(\frac{\pi (N+1) x}{L}\right)}{\sin\left(\frac{\pi x}{L}\right)},
\eea
where $L$ is the physical size of the box.
To analyze the interacting density profiles, we rewrite the above as
\bea
\label{Eq:SineTransf}
\frac{L}{N\!+\!1} n(x/L)\sin\left(\frac{\pi x}{L}\right) &=& \sin\left(\frac{\pi x}{L}\right) \!-\! \frac{1}{N\!+\!1}\sin\left(\frac{\pi (N\!+\!1) x}{L}\right) \nonumber \\
&=& \sum_{k=1}^{N^{}_x} B^{}_k \sin\left(\frac{\pi k x}{L}\right) \equiv \tilde n(x/L),
\eea
where we have used a general Fourier expression obeying hard-wall boundary conditions in the last equality.
Obviously, the noninteracting system satisfies $B^{}_k=0$ for all $k$ except the cases of $B^{}_1 = 1$ and $B^{}_{N+1} = -1/(N+1)$.
Note that the multiplication by $\sin\left(\frac{\pi x}{L}\right)$ necessarily obliterates any information at $x=0$, but we already
know that the density vanishes at that point.

In Fig.~\ref{Fig:SineTransf} we show the coefficients $B^{}_k$ as a function of $k$ and the coupling strength, for a representative case.
The dominant contributions are always at $k=1$ and $k=N+1$, regardless of the strength of the coupling.
This is a remarkably clean signal that can be experimentally verified and which is especially surprising in light of bosonization analyses.
The latter indicate that the denominator $\sin({\pi x}/{L})$ should appear elevated to a non-trivial power (i.e. different from unity) 
that is directly related to the parameters of the low-energy effective theory (see e.g. Ref.~\cite{Eggert}).

The physical origin of the $k=1$ and $k=N+1$ peaks can be gleaned from the non-interacting result Eq.~(\ref{Eq:nxNonInt}).
The former provides the ``overall'' or ``average'' density in the bulk and is responsible for (all but one of) the particle number count; 
it therefore survives in the thermodynamic limit. The second term (and presumably every other non-vanishing term in the interacting
case) is associated with the physics at the Fermi surface.

\subsection{One-body density matrix}
%
In order to characterize the one-body density matrix in an efficient way, we perform a more general version of the spectral analysis previously applied to the density profile.
The non-interacting one-body density matrix may be computed trivially by collecting contributions from the $N/2$ occupied orbitals:
\bea
\label{Eq:G1}
G_{1}(x,x')
&=&\frac{2}{L}\sum_{k=1}^{N/2}\sin\left(\frac{\pi k x}{L}\right)\sin\left(\frac{\pi k x'}{L}\right)  \\
&=&\frac{1}{2L}\left[\frac{\sin\left(\frac{\pi (N+1) (x-x')}{2L}\right)}{\sin\left(\frac{\pi(x-x')}{2L}\right)}-\frac{\sin\left(\frac{\pi (N+1) (x+x')}{2L}\right)}{\sin\left(\frac{\pi(x+x')}{2L}\right)}\right] \nonumber.
\eea
Again, we rewrite the above as (for $x\ne x'$)
%
%
\bea
\label{Eq:G1Transf}
&& \!\!\!\!\!\!\!\! 2L G_{1}(x,x')\sin\left(\frac{\pi(x-x')}{2L}\right)\sin\left(\frac{\pi(x+x')}{2L}\right)\nonumber \\
&&\ \ \ = \sin\left(\frac{\pi Nx}{2L}\right)\sin\left(\frac{\pi(N/2+1)x'}{L}\right) \nonumber \\
&& \ \ \ \ \ \ \ - \sin\left(\frac{\pi (N/2+1)x}{L}\right)\sin\left(\frac{\pi Nx'}{2L}\right) \nonumber \\
&&\ \ \ = \sum_{k=1}^{N_x} \sum_{k'=1}^{N_x} B_{kk'} \sin\left(\frac{\pi k x}{L}\right) \sin\left(\frac{\pi k' x'}{L}\right) \nonumber \\ 
&&\ \ \ \equiv \tilde G_{1}(x,x'),
\eea
%
%
where we see that $B_{kk'}=0$ for all $k$, $k'$ except for the cases of $B_{\frac{N}{2},\frac{N}{2}+1} = -B_{\frac{N}{2}+1,\frac{N}{2}} = 1$, corresponding to $k_F$ and the first excited state above
the Fermi level. Note that the antisymmetry in the coefficients was introduced in Eq.~(\ref{Eq:G1Transf}) via the antisymmetry in $x$ and $x'$. 

\begin{figure}[t]
\includegraphics[width=1.0\columnwidth]{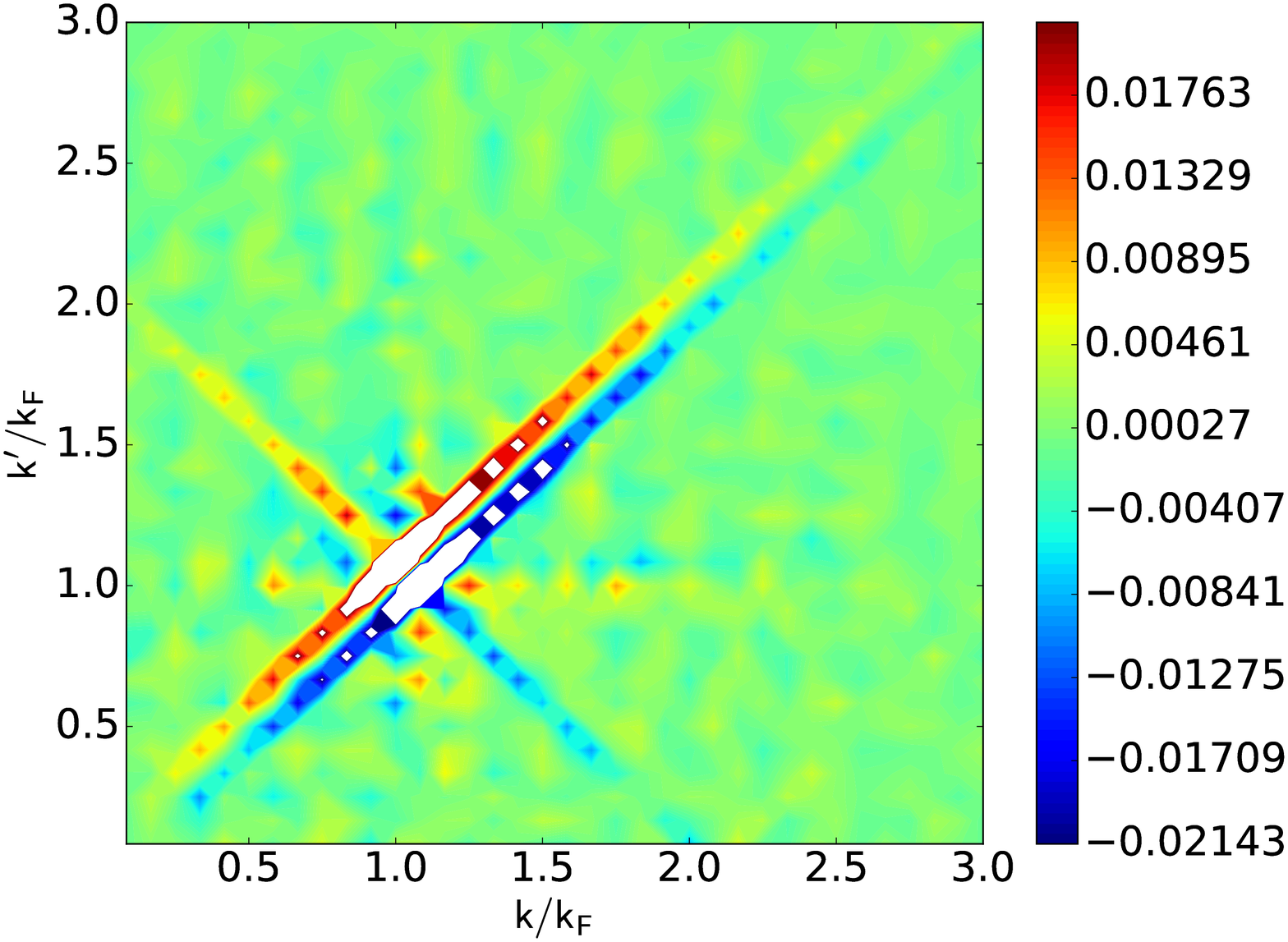}
\includegraphics[width=1.0\columnwidth]{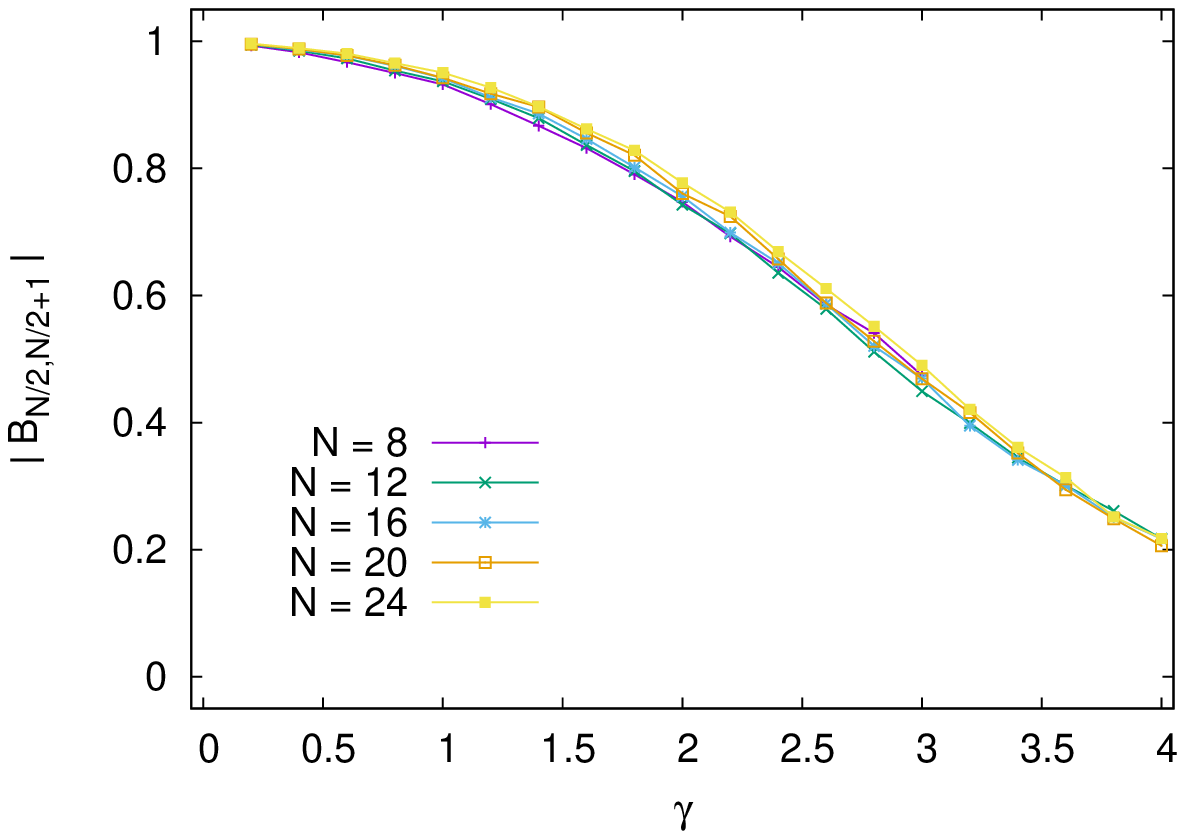}
\caption{\label{Fig:OBDM} Top: Hard-wall transform coefficients $B_{k,k'}$ of Eq.~(\ref{Eq:G1Transf}), 
as a function of $k = L p_k/\pi$ and $k' = L p_{k'}/\pi$, for $N_x^{}=80$, $\gamma=4.0$, and $N=24$.
Bottom: Amplitude variation of the main peak $B_{\frac{N}{2},\frac{N}{2}+1}$ as a function of $\gamma$
for several particle numbers.
}
\end{figure}

Interestingly, the non-zero $B_{kk'}$ in the non-interacting case display the property that $|k-k'|=1$ and $k+k'=N+1$, reflecting the 
same wavelengths observed in the spatial density profile, $n(x)$. As seen in Fig.~\ref{Fig:OBDM} (top), in the presence of attractive 
interactions, amplitude from $B_{\frac{N}{2},\frac{N}{2}+1}$ and its antisymmetric partner appears to shift
to neighboring modes, which continue to obey the same relations for $|k-k'|$ and $k+k'$. This decay of the main peak is enhanced as 
$\gamma$ is increased, as Fig.~\ref{Fig:OBDM} (bottom) demonstrates.

%
%

\subsection{Quasi-momentum distribution}

In translationally invariant systems, the shape of the momentum distribution can 
reveal several aspects of a many-body system, ranging from condensation to short-range 
correlations and spatial structure.
In the present case, the hard-wall boundaries break translation invariance, such that the 
eigenstates of the system in the absence of interactions are given by the usual standing waves 
with quantized momentum $p_k = k\pi/L$, as mentioned above.
In this section we present the occupation distribution in the single-particle space of those
standing waves; we will refer to that distribution as the quasi-momentum distribution $n(k)$.

\begin{figure}[t]
\includegraphics[width=1.0\columnwidth]{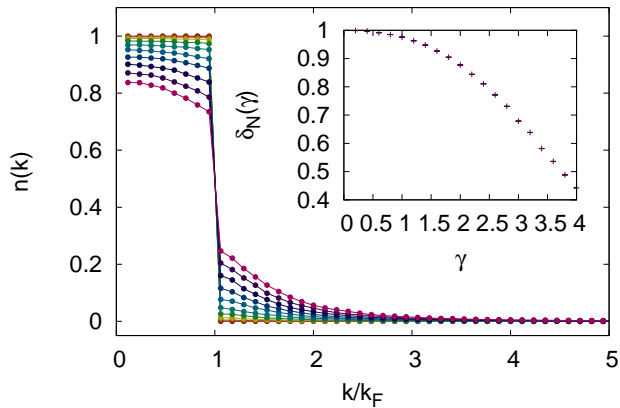}
\caption{\label{Fig:MomDist}  Quasi-momentum distribution $n(k)$ for $N=16$ particles as a function of 
$\gamma=0.2,0.6,1.0,...,3.8$ (top to bottom around $k=0$), for $N_x^{}=80$.
Inset: $\gamma$-dependence of the discontinuity $\delta_N^{}$ in $n(k)$ at the Fermi surface.
While $\delta_N^{}$ could depend on $N$,
our results for different $N$ agree within our statistical and systematic uncertainties.
}
\end{figure}

Figure~\ref{Fig:MomDist} shows $n(k)$ per spin for a representative particle number $N=16$, and for 
several couplings. For non-interacting systems, the ground-state distribution is the expected
step function, with a unit discontinuity at the Fermi surface. As attractive interactions are turned on, 
pairing correlations begin to dominate, and the system progressively becomes more
bosonic. This is clearly seen in the variation of the discontinuity $\delta^{}_N$ in $n(k)$ as a function of $\gamma$:
$\delta^{}_N$ decreases monotonically as $\gamma$ is increased.
The inset of Fig.~\ref{Fig:MomDist} shows $\delta^{}_N(\gamma)$ for $N=16$ particles.
For the couplings studied here, it appears that $\delta^{}_N(\gamma)$ decreases without 
a bound. However, it seems difficult to imagine a situation in which $\delta^{}_N(\gamma) < 0$ and the energy is minimized. 
Under the assumption that that does not happen, we expect that
$\delta^{}_N(\gamma)$ approaches a non-negative constant asymptotically at large $\gamma$, i.e. the curve described
by the data in the inset of Fig.~\ref{Fig:MomDist} should eventually change convexity as it approaches a
constant from above. To determine whether this conjecture holds, however, calculations in the
strong-coupling, hard-core boson limit are needed~\cite{SCLimit}.

%
%


\subsection{Tan's contact density and contact}

Short-range interactions, like the one studied here, induce correlations whose short-distance (high-momentum) form
is encoded in Tan's contact~\cite{Tan}. Indeed, the short-distance dynamics is governed by the shape of 
the relative-coordinate wavefunction of the two-body problem, up to an overall factor that encapsulates
many-body effects as well as effects due to external trapping potentials. The latter is the so-called contact $C$.
One of the many ways to define the contact is through the Feynman-Hellmann theorem as applied to the
variation of the ground-state energy with respect to the scattering length, which yields
\beq
C = - g \langle \hat V\rangle,
\eeq
where $\langle \hat V\rangle$ is the ground-state expectation value of the interaction energy.
Equation~(\ref{Eq:InteractionV}) indicates that $C$ is given by an integral over the on-site correlation function among different spins, 
which for brevity we will call ``contact density''. In our case, the presence of hard walls yields a spatially varying contact density $\mathcal C(x/L)$.
Note that since $g$ has units of inverse length (i.e. momentum), $C$ has dimensions of inverse length cubed. Moreover, $C$ is an extensive quantity.
Therefore, below we use the intensive dimensionless form $C/ (N k_F^3)$; for the contact density we use $\mathcal C(x/L) / k_F^4$.
\begin{figure}[t]
\includegraphics[width=1.0\columnwidth]{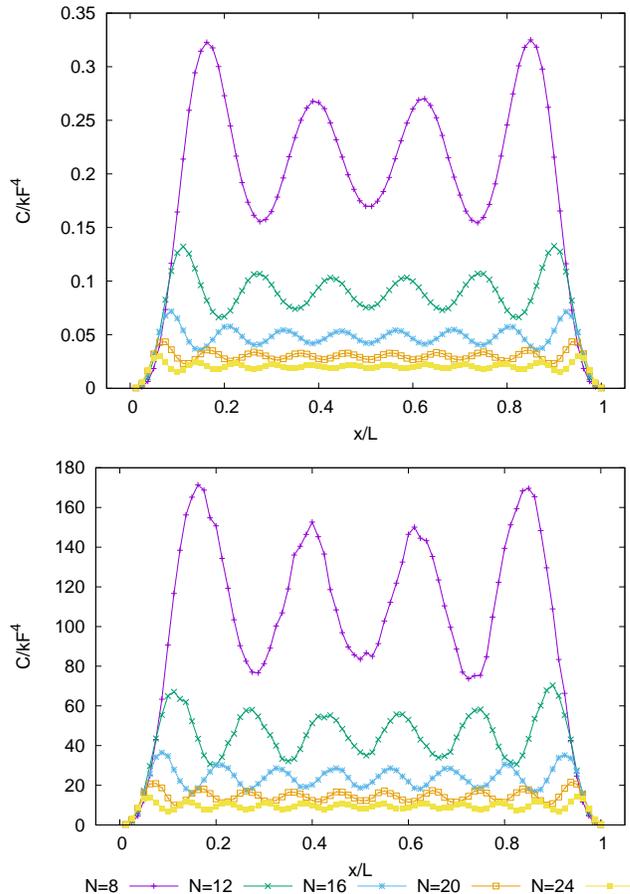}
\caption{\label{Fig:ContactDensity}  Contact density $\mathcal C(x/L)$ as a function of the scaled position $x/L$
for $N^{}_{x}=80$ at weak coupling ($\gamma=0.2$, top panel) and strong coupling ($\gamma=3.0$, bottom panel), 
for particle numbers $N=8,12,16,20,24$ (top to bottom).
}
\end{figure}

\begin{figure}[t]
\includegraphics[width=1.0\columnwidth]{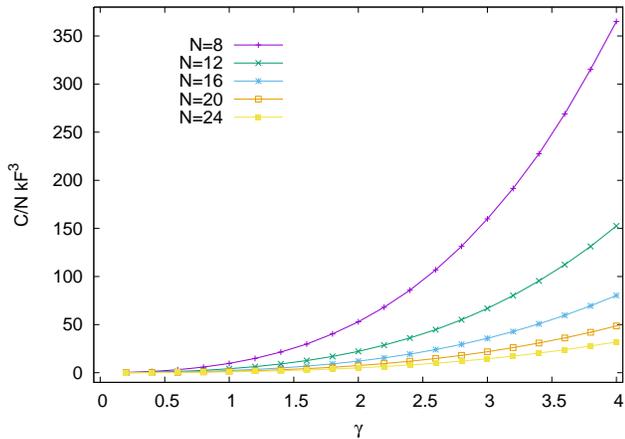}
\caption{\label{Fig:ContactGS} 
Tan's contact of $N=8,...,24$ unpolarized fermions in a 1D segment (discretized using $N_x^{} = 80$ points), 
in units of its non-interacting counterpart $E_\text{FG}^{}$, as a function of the dimensionless coupling $\gamma$.
}
\end{figure}

In Fig.~\ref{Fig:ContactDensity} we show $\mathcal C(x/L)$ as a function of the scaled position $x/L$ and in units
of $k_F^4$ at two different couplings (top and bottom). In terms of number of peaks and valleys, the oscillations in 
$\mathcal C(x/L)$ follow the same pattern as those of the density $n(x/L)$. The amplitude variations are such that
the minima are at roughly one half of the value of the maxima, regardless of coupling or particle number. While that
trend remains as $N$ is varied, the average value of $\mathcal C(x/L)/k_F^4$ away from the boundaries does 
seem to quickly approach a limit as $N$ is increased, and the overall amplitude of the oscillations also decreases
markedly. The integrated contact $C/ (N k_F^3)$, shown in Fig.~\ref{Fig:ContactGS}, also seems to approach the 
thermodynamic limit very quickly. This finding is in line with the observations of Ref.~\cite{GSC1D}, which studied the
few- to many-body progression of one-dimensional fermions with periodic boundary conditions.

%

\section{Summary and Conclusions}

Studying systems with hard-wall boundaries via ultracold atoms is a way to access
the behavior of the uniform-space limit, but when the boundaries are explicitly accounted for, 
it is also a different way to probe strongly coupled matter.
In this work, we set out to characterize an elementary yet non-trivial many-body problem in that situation:
spin-1/2 fermions with a zero-range interaction in a one-dimensional hard-wall box.
We computed, in a fully ab initio fashion, the ground-state energy, density profiles, momentum distribution, and
Tan's contact density. Together, these quantities provide a basic yet comprehensive understanding of the
effects of the boundaries and how they disappear in the thermodynamic limit. In short, we find that, upon scaling
the density and the contact density by appropriate powers of $k_F$, the large-$N$ limit is approached surprisingly quickly,
as the largest changes happen in the regime we studied, where $N=8$ -- $24$.

The hard-wall boundaries lead to oscillating density profiles; such Friedel oscillations are characteristic
not only of systems with interfaces but also of situations where an impurity is present.
Our work clarifies the quantitative changes in the oscillation pattern due to short-range attractive interactions,
which induce pairing correlations.
We find that those effects are captured by a parametrization of the density which, while based on the 
noninteracting result, provides a remarkably clean way to understand the behavior of the density
even in strongly coupled situations. We extended that analysis to the one-body density matrix, and further 
complemented it by computing the quasi-momentum distribution. The latter shows clear interaction effects as 
a change in the discontinuity at the Fermi surface.

Aside from the above, we calculated the ground-state energy (which we provide as a benchmark for other approaches 
and future studies) and the Tan contact density (i.e. on-site pairing correlations) which encode all of the short-distance
behavior of response functions via the operator product expansion.

We carried out this work by discretizing space and (imaginary) time and using the auxiliary-field path-integral
representation of the many-body problem. We estimated that path integral using stochastic methods, namely the hybrid 
Monte Carlo algorithm. 
Our study used lattices of up to $N_x^{}=80$ points and covered weakly to strongly interacting regimes in systems of 
up to $N=24$ particles. This Monte Carlo approach is one of the few tools that, in one dimension, can provide fully
non-perturbative and well-controlled access to the physics of strongly coupled matter.

Our work is a first step towards: a) exploring the detailed structure of two-body correlations, b) studying finite-temperature 
effects, and c) higher dimensions, including mixed-dimensions. Point b) is particularly important because finite-T 
effects are only cleanly accessible with exact diagonalization or with QMC, as the Bethe ansatz does not provide a controlled 
approximation in that case (at least not for all temperatures). 

\acknowledgments
{
We gratefully acknowledge discussions with D. Lee, T. Sch\"afer, and J. E. Thomas (NCSU), 
as well as with L. Rammelm\"uller (TU Wien and TU Darmstadt).  
This material is based upon work supported by the 
National Science Foundation under Grants No. 
PHY1306520 (Nuclear Theory Program) and
PHY1452635 (Computational Physics Program).
}



\end{document}